  \documentclass[12pt]{article}
  \usepackage{amssymb}
  \usepackage{latexsym}

 \renewcommand{\theequation}{\arabic{section}.\arabic{equation}}
 \catcode`\@=11 \@addtoreset{equation}{section}\catcode`\@=12

 \def\barr{\left(\begin{array}}
 \def\earr{\end{array}\right)}

\newcommand{\R}{ {\mathbb R} }
\newcommand{\eps}{ \varepsilon }

\newcommand{\p}{\partial}
\newcommand{\fnm}{\footnotemark}
\newcommand{\fnt}{\footnotetext}

\begin{document}

 \begin{center}

 \large \bf On cosmological-type  solutions
            in  multi-dimensional  model with  Gauss-Bonnet  term
           \end{center}

 \vspace{0.3truecm}

 \begin{center}

 \normalsize\bf V. D. Ivashchuk\fnm[1]\fnt[1]{e-mail:
  ivashchuk@mail.ru},

\vspace{0.3truecm}

 \it Center for Gravitation and Fundamental Metrology,
 VNIIMS, 46 Ozyornaya ul., Moscow 119361, Russia

 \it Institute of Gravitation and Cosmology,
 Peoples' Friendship University of Russia,
 6 Miklukho-Maklaya ul., Moscow 117198, Russia

\end{center}

 \begin{abstract}

  A $(n +1)$-dimensional  Einstein-Gauss-Bonnet (EGB) model is
  considered. For  diagonal  cosmological-type metrics,
  the equations of motion are reduced to a set of Lagrange equations.
  The effective  Lagrangian contains  two ``minisuperspace'' metrics on
  $\R^{n}$. The first one is the well-known 2-metric
  of  pseudo-Euclidean signature  and the second one is the Finslerian
  4-metric that is proportional to   $n$-dimensional
  Berwald-Moor 4-metric.   When  a ``synchronous-like'' time
  gauge is considered   the equations of motion are reduced to
  an autonomous system of first-order differential equations.
  For the case of the ``pure'' Gauss-Bonnet model, two
   exact solutions with power-law and exponential
  dependence of scale factors (with respect to ``synchronous-like'' variable)
  are obtained. (In the cosmological case the power-law solution was  considered
  earlier in papers of N. Deruelle,  A. Toporensky, P. Tretyakov
  and S. Pavluchenko.)   A  generalization of the effective Lagrangian
  to the Lowelock  case is conjectured. This hypothesis  implies
  existence of exact solutions with power-law
   and exponential  dependence of scale factors
   for the ``pure'' Lowelock model of $m$-th  order.

 \end{abstract}

  \newpage

\section{Introduction}

Here we deal with $D$-dimensional gravitational model with the
Gauss-Bonnet term. The action reads
\begin{equation}
 S =  \int_{M} d^{D}z \sqrt{|g|} \{ \alpha_1 R[g] +
             \alpha_2  {\cal L}_2[g] \},
   \label{1.1}
 \end{equation}
where $g = g_{MN} dz^{M} \otimes dz^{N}$ is the metric defined on
the manifold $M$, ${\dim M} = D$, $|g| = |\det (g_{MN})|$ and

\begin{equation}
  {\cal L}_2 = R_{MNPQ} R^{MNPQ} - 4 R_{MN} R^{MN} +R^2
   \label{1.2}
 \end{equation}
is the standard Gauss-Bonnet term. Here $\alpha_1$ and $\alpha_2$
are constants.  The appearance of the renormalizable Gauss-Bonnet
term as well as quadratic Riemann curvature terms in
multidimensional gravity is motivated by string theory
 \cite{Zwiebach,GBstrings1,GBstrings2,GBstrings3,GBstrings4}.
 (For a review of fourth-order  gravity in $D=4$, see  \cite{Schmidt}.)

 At present, the so-called Einstein-Gauss-Bonnet (EGB) gravity and
 its modifications are intensively used in  cosmology,
 see \cite{NojOd0,CElOdZ} (for $D =4$),
 \cite{Ishihara,ElMakObOsFil,BambaGuoOhta,TT,KirMPTop,PTop,KirMak}
 and references therein,  e.g. for explanation  of  accelerating
 expansion of the Universe following from
 supernovae (type Ia) observational data \cite{Kowalski}.
 Certain exact solutions in multidimesional EGB cosmology
 were obtained in  \cite{Ishihara}-\cite{KirMak} and
 some other papers.

 EGB gravity is also  intensively investigated  in a context of
 black-hole physics. The  most  important results here are related
 with  the well-known Boulware-Deser-Wheeler solution
 (corresponding to
 the Schwarzschild-Tangherlini solution in general relativity)
 \cite{BoulDes,Wheel} and its generalizations
 \cite{Wheel2,Wilt,Cai,CvetNojOd},  for a review
 and references, see \cite{GarGir,Charm}.
 For certain applications of  brane-world models with
 Gauss-Bonnet term, see review \cite{BrKonMel} and references therein.

 Here we are interested in the cosmological (type) solutions with diagonal
 metrics (of Bianchi-I-like type) governed by scale factors
 depending upon one variable.

  For  $\alpha_2 = 0$, we have
 the Kasner type solution with the metric
  \begin{equation}
  g= - d \tau \otimes d \tau  +
  \sum_{i=1}^{n}  A_i^2 \tau^{2p^i} dy^i \otimes dy^i, \label{1.3}
  \end{equation}
  where  $A_i > 0$ are arbitrary constants, $D = n +1$   and parameters
  $p^i$ obey the relations
     \begin{eqnarray}
      \sum_{i=1}^n  p^i = 1,  \label{1.4}\\
      \sum_{i=1}^n  (p^i)^2  = 1  \label{1.5}
      \end{eqnarray}
 and hence
        \begin{equation}
        \sum_{ 1 \leq i < j \leq n} p^i p^j =
       \frac{1}{2} ( \sum_{i=1}^n  p^i )^2 -
       \frac{1}{2} \sum_{i=1}^n  (p^i)^2
         = 0 .   \label{1.5a}
        \end{equation}
 For $D =4$, this is  the well-known Kasner solution \cite{Kasner}.
 The set of eqs.  (\ref{1.4}), (\ref{1.5})
  is equivalent to the set of eqs. (\ref{1.4}), (\ref{1.5a}).

  In \cite{Deruelle}, a Einstein-Gauss-Bonnet (EGB) cosmological
 model  was  considered. For ``pure''
 Gauss-Bonnet (GB) case $\alpha_1 = 0$ and
 $\alpha_2 \neq 0$, N. Deruelle has obtained a cosmological
 solution with the metric (\ref{1.3}) for $n = 4, 5$  and parameters obeying
 the relations
      \begin{eqnarray}
      \sum_{i=1}^n  p^i = 3,  \label{1.6}\\
      \sum_{1 \leq i < j < k < l \leq n} p^i p^j p^k p^l = 0  \label{1.7}.
      \end{eqnarray}
  It was  reported by A. Toporensky and P. Tretyakov
  in \cite{TT} that this solution was  verified
  by them for $n = 6,7$. In recent paper by S. Pavluchenko \cite{Pavl}
  the power-law solution was verified for all $n$ (and also generalized
  to the Lowelock case  \cite{Low}).

 In this paper we give a  derivation of  the ``power-law''
 (cosmological type) solution   for arbitrary $n$.
 We also show that for $D \neq 4 $ this solution in ``pure''
 GB cosmology is unique in a class of solutions with power-law
 dependence  of scale  factors: $a_i(\tau) = A_i \tau^{p^i}$, when
 the parameters $p^1,...,p^n$ contain more than two non-zero numbers.
 When $(n-2)$ parameters among $p^i$ are zero, say $p^3 =...= p^n =0$,
 than the metric (\ref{1.3}) obeys the equations of motion (for $\alpha_1 =
 0$) for arbitrary values of two Kasner-like parameters (say $p^1, p^2$).

  The numerical analysis
  of cosmological solutions in EGB gravity for $D =  5, 6$ \cite{PTop}
   shows  that the  singular   ``power-law'' solutions
   (\ref{1.3}), (\ref{1.6}), (\ref{1.7})
   (e.g. with a little generalization of scale factors $a_i(\tau) = A_i (\tau_0 \pm
  \tau)^{p^i}$, where $\tau_0$ is constant) appear as asymptotical solutions for certain
  initial values as well as Kasner-type solutions
  (\ref{1.3})-(\ref{1.5}) do.

 The paper is organized as follows. In Section 2
 the equations of motion for
  $(n +1)$-dimensional EGB model are considered.
  For  diagonal  cosmological type metrics
  the equations of motion are reduced to a set of Lagrange equations
  corresponding to certain ``effective'' Lagrangian
  (in agreement with \cite{Deruelle,Pavl} for cosmological case).
  Section 3 is devoted to  the case of the ``pure'' Gauss-Bonnet model.
  Two exact solutions: with power-law and exponential
  dependence of scale factors (with respect to ``synchronous-like'' variable)
  are obtained.
  In Section 4   the equations of motion are reduced to
  an autonomous system of first order differential equations
  (when  a ``synchronous-like'' time  gauge is considered).
   For $\alpha_1 \neq 0$ and $\alpha_2 \neq 0$
   it is shown that for any non-trivial  solution with
  exponential dependence of scale factors $a_i(\tau) = A_i \exp( v^i
  \tau)$, $i = 1,...,n$, there are no more than
  three different  numbers among  $v^1,...,v^n$.
  In Section 5 a  generalization of the effective Lagrangian
  to the Lowelock  case is conjectured and
   exact solutions with power-law
   and exponential  dependence of scale factors
   for the ``pure'' Lowelock model of $m$-th order are presented.
   (See also \cite{Deruelle,Pavl} for ``power law''  cosmological solutions.)
   Certain useful relations and proofs are collected in Appendix.

\section{The cosmological type model and its effective Lagrangian}

 \subsection{The set-up }

 Here we consider the manifold
 \begin{equation}
   M = \R_{*}  \times M_{1} \times \ldots \times M_{n}, \label{2.1}
 \end{equation}
 with the metric
 \begin{equation}
  g= w e^{2{\gamma}(u)} du \otimes du  +
 \sum_{i=1}^{n} e^{2\beta^i(u)} \eps_{i} dy^i \otimes dy^i, \label{2.2}
 \end{equation}
 where $w = \pm 1$ and any $M_i$ is 1-dimensional manifold with
 the metric $g^i = \eps_{i} dy^i \otimes dy^i$,
 $ \eps_{i}= \pm 1$, $i = 1, \dots, n$.
 Here and in what follows
  $\R_{*} = (u_{-},u_{+})$ is an open subset in $\R$.
  (Here we identify $g^i$ with $\hat{g}^{i} = p_{i}^{*} g^{i}$ which is the
   pullback of the metric $g^{i}$  to the manifold  $M$ by the
  canonical projection: $p_{i} : M \rightarrow  M_{i}$,
  $i = 1,\ldots, n$.)
 The functions ${\gamma}(u)$ and
 $\beta^i(u)$,  $i = 1,\ldots, n$, are smooth on
 $\R_{*} = (u_{-},u_{+})$.

 For $w =  -1$, $ \eps_{1}=  ... = \eps_{n} = 1$
 the metric (\ref{2.2}) is a cosmological one while
 for $w =  1$, $ \eps_{1}= -1$,
 $\eps_{2} = ... =\eps_{n} = 1$ it describes static
 configurations.

 According to Appendix A, the integrand in  (\ref{1.1}), when the
 metric (\ref{2.2}) is substituted, reads as follows
   \begin{equation}
    \sqrt{|g|} \{ \alpha_1 R[g] +
               \alpha_2  {\cal L}_2[g] \} = L + \frac{df}{du},
    \label{2.3}
   \end{equation}
 where
    \begin{eqnarray}
    L = \alpha_1 L_1 +  \alpha_2 L_2,
    \label{2.4}\\
      L_1 = (-w) e^{-\gamma + \gamma_0} G_{ij} \dot{\beta}^i
      \dot{\beta}^j,
         \label{2.5}   \\
       L_2  = - \frac{1}{3}  e^{- 3 \gamma + \gamma_0}
         G_{ijkl} \dot{\beta}^i \dot{\beta}^j \dot{\beta}^k
         \dot{\beta}^l,
              \label{2.6}
    \end{eqnarray}
  $\gamma_0 = \sum_{i =1}^{n} \beta^i$ and
  \begin{eqnarray}
       G_{ij} = \delta_{ij} -1,
         \label{2.10}   \\
       G_{ijkl}  = (\delta_{ij} -1)(\delta_{ik} -1)(\delta_{il} -1)
       (\delta_{jk} -1)(\delta_{jl} -1)(\delta_{kl} -1)
       \label{2.11}
      \end{eqnarray}
      are respectively the components of two ``minisuperspace'' metrics on
      $\R^{n}$. (For cosmological case see  also \cite{Deruelle,Pavl,Iv-09}.)
      The first one is the well-known 2-metric
      of  pseudo-Euclidean signature: $<v_1,v_2> = G_{ij}v^i_1 v^j_2$
      and the second one is the Finslerian 4-metric:
      $<v_1,v_2,v_3,v_4> = G_{ijkl}v^i_1 v^j_2 v^k_3 v^l_4$,
      $v_s = (v^i_s) \in \R^n$,
      where $<.,.>$ and $<.,.,.,.>$ are respectively
      $2$- and $4$-linear symmetric forms on $\R^n$.
     (Here we denote $\dot{A} = dA/du$ etc.)

    In (\ref{2.3}) the
    function $f = f(\gamma, \beta, \dot{\beta})$ has the following
    form:
     \begin{equation}
     f = \alpha_1 f_1 +  \alpha_2 f_2,
     \label{2.7}
     \end{equation}
    where $f_1$ and $f_2$ are defined in Appendix A (see
    (\ref{A.18f1}) and  (\ref{A.18f2})).

      The derivation of (\ref{2.4})-(\ref{2.6}) is based on
      the  relations   obtained in  Appendix A
      (see (\ref{A.18L1}),  (\ref{A.18L2}))
        and the following identities
      \begin{eqnarray}
       G_{ij}v^i v^j = \sum_{i =1}^{n} (v^i)^2 -
        (\sum_{i =1}^{n} v^i )^2,
         \label{2.12}   \\
       G_{ijkl}v^i v^j v^k v^l  = (\sum_{i =1}^{n} v^i )^4
        - 6 (\sum_{i =1}^{n} v^i )^2
        \sum_{j =1}^{n} (v^j)^2
         \nonumber \\
        +     3 ( \sum_{i =1}^{n} (v^i)^2 )^2
        + 8  (\sum_{i =1}^{n} v^i )
        \sum_{j =1}^{n} (v^j)^3  - 6 \sum_{i =1}^{n} (v^i)^4.
       \label{2.13}
      \end{eqnarray}
       The first identity  (\ref{2.12}) is a trivial one.
       The second one (\ref{2.13}) may be verified by
        straightforward calculations (see Appendix B).

      It follows immediately from the definitions
      (\ref{2.10}) and (\ref{2.11}) that
      \begin{eqnarray}
       G_{ij}v^i v^j = -2 \sum_{i < j} v^i v^j,
         \label{2.14}   \\
       G_{ijkl}v^i v^j v^k v^l  = 24 \sum_{i < j < k < l} v^i v^j
       v^k v^l .
       \label{2.15}
      \end{eqnarray}

       Due to (\ref{2.15}),  $G_{ijkl}v^i v^j v^k
       v^l$ is zero for $n = 1, 2, 3$ ($D = 2, 3, 4$).
       For $n = 4$ ($D = 5$),  $G_{ijkl}v^i v^j v^k v^l = 24 v^1 v^2 v^3
       v^4$ and our  4-metric is proportional to the well-known
       Berwald-Moor 4-metric \cite{Berwald,Moor}
       (see also \cite{Bogos,GarPav} and references therein).
       We remind the reader that
       the 4-dimensional Berwald-Moor 4-metric obeys the
       relation: $<v,v,v,v>_{BM} =v^1 v^2 v^3v^4$.
       The  Finslerian 4-metric with components (\ref{2.11})
       coincides up to a factor with the $n$-dimensional
       analogue  of the Berwald-Moor 4-metric.

     \subsection{The equations of motion }

    The equations of motion corresponding to the action (\ref{1.1})
  have the following form
  \begin{equation}
   {\cal E}_{MN} = \alpha_1 {\cal E}^{(1)}_{MN}
     + \alpha_2 {\cal E}^{(2)}_{MN} = 0,
   \label{1.3e}
 \end{equation}
  where
  \begin{eqnarray}
   {\cal E}^{(1)}_{MN} = R_{MN} - \frac{1}{2} R g_{MN},
   \label{1.3a} \\
   {\cal E}^{(2)}_{MN} = 2(R_{MPQS}R_N^{\ \ PQS} - 2 R_{MP} R_N^{\ \ P}
   \nonumber \\
   -2 R_{MPNQ} R^{PQ} + R R_{MN}) -  \frac{1}{2} {\cal L}_2  g_{MN}.
   \label{1.3b}
   \end{eqnarray}

    The field equations (\ref{1.3e}) for the metric
    (\ref{2.2}) are equivalent to the Lagrange equations
    corresponding to the Lagrangian $L$ from (\ref{2.4}).
    This follows from the relations
      \begin{eqnarray}
       {\cal E}_{00}(-2w) \exp(\gamma_0 - \gamma) =
       \frac{\p L}{\p \gamma},
       \label{2.16a} \\
       {\cal E}_{ii}(-2 \eps_{i})
       \exp(\gamma + \gamma_0 - 2\beta^i)  =
       \frac{\p L}{\p \beta^i} - \frac{d}{du} \frac{\p L}{\p \dot{\beta}^i}
       \label{2.16b}, \\
       {\cal E}_{0i} = 0,
        \label{2.16c}
        \end{eqnarray}
         $i = 1,\ldots, n$.

    Formulas (\ref{2.16a})-(\ref{2.16c}) may be verified
    just by  straightforward calculations based on the relations for the
    Riemann tensor from Appendix A. But there exists a more
    ``economic'' way to prove these formulas using:
    {\bf (i)} the diagonality  of the matrix ${\cal E}_{MN}$
    (in coordinates $(y^M) = (y^0 =u,  y^i)$);
    {\bf (ii)} the dependence of this matrix only on one variable
    $u$, i.e.   ${\cal E}_{MN} = {\cal E}_{MN}(u)$;
    {\bf (iii)}  the relation  (\ref{2.3}).
      The proof of (\ref{2.16a})-(\ref{2.16c}) is given in  Appendix C.

    Thus, equations (\ref{1.3e}) read as follows
    \begin{eqnarray}
       w \alpha_1  G_{ij} \dot{\beta}^i \dot{\beta}^j
        + \alpha_2  e^{- 2 \gamma}
         G_{ijkl} \dot{\beta}^i \dot{\beta}^j \dot{\beta}^k
         \dot{\beta}^l = 0,  \label{2.17} \\
         \frac{d}{du}[ - 2w \alpha_1  G_{ij} e^{-\gamma +
         \gamma_0}
          \dot{\beta}^j  \qquad \qquad   \nonumber \\
        -  \frac{4}{3} \alpha_2 e^{- 3 \gamma + \gamma_0}
         G_{ijkl}  \dot{\beta}^j \dot{\beta}^k
         \dot{\beta}^l] - L = 0,   \label{2.18}
      \end{eqnarray}
     $i = 1,\ldots, n$. Due to (\ref{2.17})
         \begin{equation}
      L=  -w \frac{2}{3} e^{-\gamma + \gamma_0}
      \alpha_1  G_{ij} \dot{\beta}^i  \dot{\beta}^j.
     \label{2.18a}
       \end{equation}

    \section{Exact solutions in  Gauss-Bonnet  model}

    Now we put $\alpha_1 = 0$ and $\alpha_2 \neq 0$, i.e. we consider the cosmological
    type model governed by the action
    \begin{equation}
      S_2 =  \alpha_2 \int_{M} d^{D}z \sqrt{|g|} {\cal L}_2[g].
    \label{3.1}
    \end{equation}

    The equations of motion (\ref{1.3e}) in this
    case read
    \begin{equation}
   {\cal E}^{(2)}_{MN} = {\cal R}^{(2)}_{MN} -
    \frac{1}{2} {\cal L}_2  g_{MN} = 0,
   \label{3.1a}
   \end{equation}
    where
    \begin{eqnarray}
   {\cal R}^{(2)}_{MN} =
   2(R_{MPQS}R_N^{\ \ PQS} - 2 R_{MP} R_N^{\ \ P}
   \nonumber \\
   -2 R_{MPNQ} R^{PQ} + R R_{MN}).
   \label{3.1b}
   \end{eqnarray}

   Due to  identity $g^{MN} {\cal R}^{(2)}_{MN} = 2 {\cal L}_2$,
    the set of eqs.  (\ref{3.1a}) for $D \neq 4$
    implies
    \begin{equation}
   {\cal L}_2  = 0.
   \label{3.1d}
   \end{equation}
   It is obvious  that the set of eqs.  (\ref{3.1a})
   is  equivalent for $D \neq 4$  to the following set of equations
   \begin{equation}
   {\cal R}^{(2)}_{MN} = 0.
   \label{3.1c}
   \end{equation}

      Equations of motion (\ref{2.17}) and (\ref{2.18})
   in this case read as follows
   \begin{eqnarray}
       G_{ijkl} \dot{\beta}^i \dot{\beta}^j \dot{\beta}^k
         \dot{\beta}^l = 0,  \label{3.2} \\
         \frac{d}{du} \left[ e^{- 3 \gamma + \gamma_0}
         G_{ijkl}  \dot{\beta}^j \dot{\beta}^k
         \dot{\beta}^l \right] = 0,   \label{3.3}
      \end{eqnarray}
     $i = 1,\ldots, n$. Here  $L = 0$ due to  (\ref{3.2}).

     Let us put $\ddot{\beta}^i = 0$ for all $i$ or,
     equivalently,
     \begin{equation}
     \beta^i = c^i u + c^i_0,
     \label{3.4}
    \end{equation}
    where $c^i$ and $c^i_0$ are constants, $i = 1,\ldots, n$.
     We also put
    \begin{equation}
      3 \gamma = \gamma_0 = \sum_{i =1}^{n} \beta^i,
     \label{3.5}
    \end{equation}
     i.e. a modified ``harmonic'' variable is used.
     Recall that in the case $\alpha_1 \neq 0$ and $\alpha_2 = 0$,
     the choice $\gamma = \gamma_0$ corresponds to the harmonic
     variable $u$ \cite{IM-top}.

     Then eqs. (\ref{3.3}) are satisfied identically
     and eq. (\ref{3.2}) gives us the following
     constraint
      \begin{equation}
      G_{ijkl} c^i c^j c^k c^l = 24 \sum_{i < j < k < l} c^i c^j
       c^k c^l = 0   \label{3.6}.
      \end{equation}

      Thus, we have obtained a class of exact cosmological type
      solutions for the Gauss-Bonnet model (\ref{3.1}) that
      is given by the metric (\ref{2.2}) with the functions
      $\beta^i(u)$ and $\gamma(u)$ from (\ref{3.4}) and
      (\ref{3.5}), respectively, and integration constants
       $c^i$ obeying (\ref{3.6}).

   \subsection{Solution with power-law dependence of scale factors}

    Let us consider the solutions with

       \begin{equation}
       \sum_{i=1}^n  c^i  \neq  0   \label{3.7}.
       \end{equation}

   Introducing the synchronous-type variable

      \begin{equation}
       \tau = \frac{1}{c} \exp(c u + c_0)   \label{3.8},
       \end{equation}
   where
      \begin{equation}
       c = \frac{1}{3} \sum_{i=1}^n  c^i, \qquad
       c_0 = \frac{1}{3} \sum_{i=1}^n  c^i_0,
       \label{3.9}
       \end{equation}

    and defining new parameters
      \begin{equation}
       p^i =  c^i/c, \label{3.10}, \qquad
       A_i = \exp[c^i_0 + p^i (\ln c - c_0)],
           \end{equation}
  $i = 1,\ldots, n$,
   we get the ``power-law'' solution  with the metric
  \begin{equation}
  g= w d \tau \otimes d \tau  +
  \sum_{i=1}^{n} \eps_{i} A_i^2 \tau^{2p^i} dy^i \otimes dy^i, \label{3.12}
  \end{equation}
 where $w = \pm 1$, $ \eps_i= \pm 1$;
  $A_i > 0$ are arbitrary constants,   and parameters
  $p^i$ obey the relations
   \begin{eqnarray}
      \sum_{i=1}^n  p^i = 3,  \label{3.13}\\
      G_{ijkl} p^i p^j p^k p^l =
      24 \sum_{i < j < k < l} p^i p^j p^k p^l = 0,  \label{3.14}
      \end{eqnarray}
 following from (\ref{3.6}), (\ref{3.7}) and (\ref{3.10}).
 This solution is a singular one for any set of parameters $p^i$,
 see Appendix D.

  In the cosmological case when $w = -1$, $\eps_i= 1$ (for all
  $i$),   this solution was obtained  earlier in \cite{Deruelle} for  $D = 5,6$
  and verified recently in \cite{Pavl}  for all $D > 4$.

   {\bf Example 1.} Let us consider the case $D = 6$ and
    $p_i \neq 0$, $i =1, \dots, 5$.
   Relations (\ref{3.13}) and (\ref{3.14}) read in this case
   as follows
   \begin{eqnarray}
      p^1 + p^2 + p^3 + p^4 + p^5 = 3,  \label{3.13e}\\
       p^1 p^2 p^3 p^4 p^5
      \left(\frac{1}{p^1}  + \frac{1}{p^2} + \frac{1}{p^3}
       + \frac{1}{p^4} + \frac{1}{p^5} \right)  = 0.   \label{3.14e}
      \end{eqnarray}
     Let us  put $p^1  = x > 0$, $p^2 = \frac{1}{x}$, $p^3  = z > 0$,
            $p^4  = y < 0$, $p^5 = \frac{1}{y}$. Then we get
     \begin{equation}
      x +  \frac{1}{x} + z + y + \frac{1}{y} = 3, \quad
      x +  \frac{1}{x} + \frac{1}{z} + y + \frac{1}{y} = 0.  \label{3.14ee}
      \end{equation}
   Subtracting the second relation in (\ref{3.14ee})
   from the first one we obtain
   $z - \frac{1}{z} = 3$ or $z = \frac{1}{2}(3 + \sqrt{13})$ ($z > 0$).
   For any $x > 0$ there are two solutions  $y = y_{\pm}(x)
   = \frac{1}{2}(- A \pm \sqrt{A^2 -4})$,
   where $A = x +  \frac{1}{x} + \frac{1}{z} > 2$.

   {\bf Proposition 1.}
   {\em For $D \neq 4$ the metric (\ref{3.12})
   is  a solution to  equations of motion (\ref{3.1a}) if and only if the set of
   parameters $p = (p^1,...,p^n)$ either obeys the relations
   (\ref{3.13}) and (\ref{3.14}), or $p = (a,b,0,...,0), (a,0,b,0,...,0), \ldots
   $, where $a$ and $b$ are arbitrary real numbers.}

   This proposition is proved in Appendix E.
   (For cosmological solutions in dimensions $D = 5,6$ see also
   \cite{Deruelle}.)

   For $D =4$ the metric (\ref{3.12}) gives a solution to
   equations of motion (\ref{3.1a}) for any set of parameters $p^i$.

   \subsection{Solution with exponential dependence of scale factors}

    Now we consider the solution with

       \begin{equation}
       \sum_{i=1}^n  c^i  =  0   \label{3.15}.
       \end{equation}

    Introducing the synchronous-type variable

      \begin{equation}
       \tau = u \exp(c_0)   \label{3.16},
       \end{equation}
   where $c_0$ is defined in (\ref{3.9})
    and defining new parameters
      \begin{equation}
       v^i =  c^i \exp(-c_0),  \qquad
       B_i = \exp(c^i_0), \label{3.17}
       \end{equation}
  $i = 1,\ldots, n$,
   we are led to the cosmological-type solution
    with the metric
  \begin{equation}
  g= w d \tau \otimes d \tau  +
  \sum_{i=1}^{n} \eps_{i} B_i^2 e^{2v^i \tau} dy^i \otimes dy^i, \label{3.19}
  \end{equation}
 where $w = \pm 1$, $ \eps_i= \pm 1$;
  $B_i > 0$ are arbitrary constants,   and parameters
  $v^i$ obey the relations
   \begin{eqnarray}
      \sum_{i=1}^n  v^i = 0,  \label{3.20}\\
      G_{ijkl} v^i v^j v^k v^l =
      24 \sum_{i < j < k < l} v^i v^j v^k v^l = 0,   \label{3.21}
      \end{eqnarray}
 following from (\ref{3.6}), (\ref{3.15}) and (\ref{3.17}).

   {\bf Example 2.} Let  $D = 6$ and
   $v_i \neq 0$, $i =1, \dots, 5$.
   Relations (\ref{3.20}) and (\ref{3.21}) read in this case
   as follows
   \begin{eqnarray}
      v^1 + v^2 + v^3 + v^4 + v^5 = 0,  \label{3.20e}\\
       v^1 v^2 v^3 v^4 v^5
      \left(\frac{1}{v^1}  + \frac{1}{v^2} + \frac{1}{v^3}
       + \frac{1}{v^4} + \frac{1}{v^5} \right)  = 0.   \label{3.21e}
      \end{eqnarray}
     We   put $v^1  = x > 0$, $v^2 = \frac{1}{x}$, $v^3  = 1$,
            $v^4  = y < 0$, $v^5 = \frac{1}{y}$. Then we get
     \begin{equation}
      x +  \frac{1}{x} + 1 + y + \frac{1}{y} = 0,            \label{3.20ee}
     \end{equation}
   For any $x > 0$ there are two solutions  $y = y_{\pm}(x)
   = \frac{1}{2}(- B \pm \sqrt{B^2 -4})$,
   where $B = x +  \frac{1}{x} + 1 \geq 3$.

     \subsection{Some other solutions}

      The solutions to equations of motion (\ref{3.2}) and (\ref{3.3})
      are not exhausted by relations  (\ref{3.4})-(\ref{3.6}).
      We give an example of another solution for $D > 4$:
      \begin{eqnarray}
          e^{- 3 \gamma + \beta^1 + \beta^2 + \beta^3}
          \dot{\beta}^1 \dot{\beta}^2 \dot{\beta}^3 = C,
          \label{3.b1} \\
          \beta^i(u) =  \beta^i_0, \qquad i > 3,
         \label{3.b2}
      \end{eqnarray}
      where $\beta^i_0$ ($i > 3$) and $C$ are arbitrary constants.
      In terms of ``synchronous'' variable $\tau$ (obeying $d \tau =
      e^{\gamma(u)} du$) this solutions reads
      as follows
        \begin{equation}
        g= w d \tau \otimes d \tau  +
        \sum_{i=1}^{n} \eps_{i} a_i^2(\tau) dy^i \otimes dy^i, \label{3.a}
        \end{equation}
       where
        \begin{eqnarray}
         \left(\frac{d a_1}{d \tau}\right) \left(\frac{d a_2}{d\tau}\right)
         \left(\frac{d a_3}{d\tau}\right) = C, \label{3.a1} \\
          a_i(\tau) = a_i^0,  \qquad i > 3, \label{3.a2}
        \end{eqnarray}
        where $a^i_0 > 0$ ($i > 3$) and $C$ are constants.
        This solution contains a special solution with
       \begin{equation}
        a_1(\tau)= a_2(\tau)= a_3(\tau) = A \tau . \label{3.a3}
       \end{equation}

        For $C = 0$ we get a special solution
        with arbitrary (smooth) functions
        $\gamma (u)$,  $\beta^1(u)$,  $\beta^2(u)$
        and constant    $\beta^i(u) =  \beta^i_0$, for $i > 2$.
        In terms of synchronous variable this solution
        is described by the metric (\ref{3.a}) with
        \begin{equation}
        a_1(\tau), a_2(\tau) - {\rm arbitrary}, \qquad
        a_i(\tau) = a_i^0 - {\rm constant}, \quad  i > 2.
         \label{3.c}
        \end{equation}

    {\bf Remark 1.} For $D = 4$, or $n= 3$, the equations
      of motion (\ref{3.2}) and (\ref{3.3})
      are satisfied identically for arbitrary (smooth)
      functions $\beta^i(u)$ and $\gamma(u)$. This
      is in agreement with that fact that in dimension $D = 4$,
      the action (\ref{3.1}) is a topological invariant and its
      variation is identically zero.

  \section{Reduction to an autonomous system of first order differential equations}

   Now we put $\gamma = 0$,
   i.e. ``the synchronous-like'' time gauge is considered. We denote $u = \tau$.
   By introducing ``Hubble-like'' variables $h^i = \dot{\beta}^i$, we
   rewrite eqs. (\ref{2.17}) and  (\ref{2.18}) in the following
   form

  \begin{eqnarray}
       w \alpha_1  G_{ij} h^i h^j
        + \alpha_2   G_{ijkl} h^i h^j h^k h^l = 0,  \label{5.1} \\
          \left[ -2 w  \alpha_1  G_{ij} h^j
        -  \frac{4}{3} \alpha_2  G_{ijkl}  h^j h^k h^l \right] \sum_{i=1}^nh^i
        \qquad \nonumber \\
          + \frac{d}{d\tau} \left[ -2 w  \alpha_1  G_{ij} h^j
           -  \frac{4}{3} \alpha_2  G_{ijkl}  h^j h^k h^l \right]
           - L    = 0,   \label{5.2}
      \end{eqnarray}
     $i = 1,\ldots, n$, where
      \begin{equation}
       L =  -w \alpha_1 G_{ij} h^i h^j
                - \frac{1}{3} \alpha_2   G_{ijkl} h^i h^j h^k h^l.
         \label{5.1a}
       \end{equation}

     Due to (\ref{5.1}),
      \begin{equation}
       L = -  \frac{2}{3} w \alpha_1  G_{ij} h^i h^j.
        \label{5.1b}
       \end{equation}

      Thus, we are led to the
     autonomous system of the first-order differential equations on
     $h^1(\tau), ..., h^n(\tau)$.

     Here we may use the relations (\ref{2.12}), (\ref{2.13})
     and the following formulas (with $v^i = h^i$)
     \begin{eqnarray}
       G_{ij}v^j = v^i - S_1,
         \label{5.3}   \\
       G_{ijkl} v^j v^k v^l
       = S_1^3  + 2 S_3 -3 S_1 S_2  +  3 (S_2  - S_1^2)  v^i
         +  6 S_1 (v^i)^2 - 6(v^i)^3,
       \label{5.4}
      \end{eqnarray}
     $i = 1,\ldots, n$, where $S_k = S_k (v) = \sum_{i =1}^n
     (v^i)^k$.    Relation (\ref{5.4}) is derived in
     Appendix B.

      Let us consider the  fixed point of the system
       (\ref{5.1}) and (\ref{5.2}):   $h^i (\tau) = v^i$ with constant $v^i$
      corresponding to the solutions
      \begin{equation}
      \beta^i = v^i \tau +
      \beta^i_0,  \label{5.4a}
      \end{equation}
       where $\beta^i_0$ are constants, $i = 1,\ldots, n$.
      In this case we obtain the metric (\ref{3.19})
      with exponential dependence of scale  factors.
      For $\alpha_1 = 0$ we get the solution (\ref{3.19})-(\ref{3.21}).

      Now we put $\alpha_1 \neq 0$ and $\alpha_2 \neq
      0$. For the fixed point  $v = (v^i)$ we have the set  polynomial equations
          \begin{eqnarray}
         G_{ij} v^i v^j
         - \alpha_w   G_{ijkl} v^i v^j v^k v^l = 0,  \label{5.5} \\
          \left[ 2   G_{ij} v^j
        -  \frac{4}{3} \alpha_w  G_{ijkl}  v^j v^k v^l \right] \sum_{i=1}^n v^i
        -  \frac{2}{3}   G_{ij} v^i v^j  = 0,   \label{5.6}
      \end{eqnarray}
     $i = 1,\ldots, n$, where  $\alpha_w = \alpha_2(-w)/\alpha_1$.
      For $n > 3$ this is a set of forth-order polynomial
      equations.

      The trivial solution $v = (v^i) = (0, ..., 0)$ corresponds
      to  a flat metric $g$.

      For any non-trivial solution $v$ we have
      $\sum_{i=1}^n v^i  \neq 0$ (otherwise one gets
      from (\ref{5.6}) $G_{ij} v^i v^j = \sum_{i =1}^{n} (v^i)^2 -
        (\sum_{i =1}^{n} v^i)^2 = 0$ and hence $v = (0, \dots, 0)$).

      Let us consider the isotropic case $v^1 = ... = v^n = a$.
      The set of equations (\ref{5.5}) and (\ref{5.6})
      is reduced to the  equation
       \begin{equation}
        n(n -1)a^2 + \alpha_w n(n -1)(n -2)(n -3) a^4 = 0.
         \label{5.7}
        \end{equation}
         For $n = 1$,  $a$ is arbitrary and $a =0$ for $n = 2,3$.
         When $n > 3$,  the non-zero solution
         to eq. (\ref{5.7}) exists only if $\alpha_w  < 0$ and
         in this case
         \begin{equation}
         a = \pm \frac{1}{\sqrt{|\alpha_w| (n -2)(n -3)}}.
         \label{5.8}
         \end{equation}
         In cosmological case $w  = -1$, this
         solution takes place when $\alpha_2/\alpha_1 < 0$.

         Here the problem of classification of all solutions
         to eqs. (\ref{5.5}), (\ref{5.6}) for given $n$ arises.
         Some special solutions of the form $(a,...,a,b,...,b)$,
         e.g. in a context of cosmology with two factor spaces,
         for certain dimensions were considered in literature.
         See, for example,
         \cite{Ishihara,ElMakObOsFil,BambaGuoOhta,KirMak}.

          Here we outline three  properties of the solutions
          to the set of polynomial equations (\ref{5.5}),
          (\ref{5.6}).

          {\bf Proposition 2.} {\em For any solution
          $v = (v^1,...,v^n)$ to  polynomial eqs. (\ref{5.5}) and
          (\ref{5.6}):

          {\bf i)} the  vector $-v = (-v^1,...,-v^n)$
          is also a solution;

          {\bf ii)} for any permutation  $\sigma$ of the set of indices  $\{1,..., n \}$
          the  vector $v = (v^{\sigma(1)},...,v^{\sigma(n)})$
          is also a solution;

           {\bf iii)}
           there are no more than
          three different  numbers among  $v^1,...,v^n$,
          when  $v  = (v^1,...,v^n) \neq (0,...,0)$ }.

           {\bf Proof.} The first item of the proposition is trivial.
           The second one follows just from relations
          (\ref{2.12}), (\ref{2.13}), (\ref{5.3}) and (\ref{5.4}).

           Now we prove the   item {\bf iii)}.
           Let us suppose that there exists a non-trivial solution
           $v = (v^1,...,v^n)$ with more than
           three different  numbers among  $v^1,...,v^n$.
           Due to  (\ref{5.4}), (\ref{5.6}) and  $\sum_{i=1}^n v^i  \neq 0$
            any number   $v^i$ obeys the cubic equation
            $C_0 + C_1 v^i + C_2 (v^i)^2  + C_3(v^i)^3 = 0$,
            with $C_3 \neq 0$,  $i = 1,\ldots, n$, and hence at most three
             numbers among $v^i$  may be different. Thus, we are led
            to a contradiction. The proposition  is proved.

            This implies that in a future  investigations of solutions to
            eqs. (\ref{5.5}) and (\ref{5.6}) for arbitrary $n$
            we will  need a consideration
            of three non-trivial cases when
             1) $v = (a,...,a)$ (see (\ref{5.8}));
             2) $v = (a,...,a,b,...,b)$
                ($a \neq b$); and
             3) $v = (a,...,a,b,...,b,c,...,c)$
                ($a \neq b$, $b \neq c$, $a \neq c$). One may put also
            $a > 0$ due to item {\bf i)}.

  \section{The generalization to the Lowelock  model}

  The action (\ref{1.1}) is a special case of  the
  Lowelock model \cite{Low}
  \begin{equation}
 S =  \int_{M} d^{D}z \sqrt{|g|}
  \left\{  \sum_{k = 1}^{m} \alpha_k {\cal L}_k \right\},
   \label{4.1}
 \end{equation}
where $\alpha_1, ...,\alpha_m$ are constants and ${\cal L}_k$ are
defined as follows
  \begin{equation}
 {\cal L}_k = 2^{-k} \delta^{M_1...M_{2k}}_{N_1...N_{2k}}
  R_{M_1 M_2}^{\ \ \ \ \ \ N_1 N_2} ...
  R_{M_{2k -1} M_{2k}}^{\ \ \ \ \ \ \ \ N_{2k -1} N_{2k}}  ,
   \label{4.2}
   \end{equation}
 $k = 1, \dots, m$. (Usually, $m$ is chosen
 as follows: $m = m(D) = [(D-1)/2]$;
  the terms with $k > m(D)$ will not give
  contributions into equations of motion.)
   Here
 \begin{equation}
 \delta^{M_1...M_{2k}}_{N_1...N_{2k}} = \sum_{\sigma}
 \varepsilon_{\sigma}
  \delta^{M_1}_{N_{\sigma(1)}}...  \delta^{M_{2k}}_{N_{\sigma(2k)}}
  \label{4.2k}
 \end{equation}
 is a generalized Kronecker tensor, totally antisymmetric in both
 groups of indices: $M_1,...,M_{2k}$ and $N_1,...,N_{2k}$.
 In (\ref{4.2k}) a sum on all permutations  of
 the set of indices  $\{1,..., 2k \}$ is assumed. Here
  $\varepsilon_{\sigma} = \pm 1$ is the parity of the permutation $\sigma$.

  It may  be  verified
  that   ${\cal L}_1 = R[g]$ and ${\cal L}_2$  (from (\ref{4.2}))
   is coinciding with the Gauss-Bonnet term (\ref{1.2}).

  \subsection{The Lagrange approach}

  Here we suggest  the following
  conjecture: the equations of
  motion for the Lowelock action (\ref{4.1})
   when  the metric  (\ref{2.2}) is substituted
  are equivalent to the Lagrange equations corresponding
  to the Lagrangian (for cosmological case see  also \cite{Deruelle,Pavl})

   \begin{equation}
    L = \sum_{k = 1}^m \alpha_k L_k, \label{4.3}
   \end{equation}
   where
     \begin{equation}
         L_k = \mu_k  \exp[- (2k - 1) \gamma + \gamma_0]
         G_{i_1 ... i_{2k}}^{(2k)} \dot{\beta}^{i_1}
          \ldots  \dot{\beta}^{i_{2k}},
              \label{4.4}
    \end{equation}
  $\gamma_0 = \sum_{i =1}^{n} \beta^i$, $\mu_k$ are rational numbers
  ($\mu_1 = -w, \mu_2 = - 1/3$) and
  \begin{equation}
              G_{i_1 ... i_{2k}}^{(2k)}  =
              \prod_{1 \leq r < s \leq 2k }
              (\delta_{i_r i_s} -1)
                \label{4.5}
      \end{equation}
      are the components of  Finslerian $2k$-metric:
      $<v_1,...,v_{2k}>_{2k} = G_{i_1 ... i_{2k}}^{(2k)}
      v^{i_1}_1 ... v^{i_{2k}}_{2k}$, $v_s = (v^i_s) \in \R^n$,
      where  $<.,...,.>_{2k}$ is a $2k$-linear symmetric form on $\R^n$,
      $k = 1, ..., m$. Here $G_{i_1 i_2}^{(2)} = G_{i_1 i_2}$
      and  $G_{i_1 i_2 i_3 i_4}^{(4)} =  G_{i_1 i_2 i_3 i_4}$,
      see (\ref{2.10}) and (\ref{2.11}).

      \subsection{Cosmological type solutions for ``pure'' $m$-th Lowelock
                   model}

     Now we put $\alpha_1 = ... = \alpha_{m-1} =  0$
     and $\alpha_{m} \neq  0$,
     i.e. we consider the cosmological
     type model governed by the ``pure'' $m$-th Lowelock
     action
    \begin{equation}
      S_m =  \alpha_m \int_{M} d^{D}z \sqrt{|g|} {\cal L}_m [g],
    \label{4.6}
    \end{equation}
    $m= 1, 2, 3, ...$.

    It may be verified along a line as it was done in the
    Section 3 that our conjecture implies the existence
    of cosmological type solutions with the
    metrics (\ref{3.12}) and (\ref{3.19}).

    For the ``power-law'' solution with the metric

      $$g= w d \tau \otimes d \tau  +
  \sum_{i=1}^{n} \eps_{i} A_i^2 \tau^{2p^i} dy^i \otimes dy^i$$

    the parameters $p^i$  obey the
    following relations
    \begin{eqnarray}
      \sum_{i=1}^n  p^i = 2m - 1,  \label{4.7}\\
       G_{i_1 ... i_{2m}}^{(2m)} p^{i_1} ... p^{i_{2m}} =
      (2m)! \sum_{i_1 <... <i_{2m}} p^{i_1} ... p^{i_{2m}} = 0.   \label{4.8}
      \end{eqnarray}
    instead of  (\ref{3.13}) and (\ref{3.14}).
    (For cosmological solutions see also \cite{Deruelle,Pavl}.)

    For the ``exponential'' solution with the metric

     $$g= w d \tau \otimes d \tau  +
      \sum_{i=1}^{n} \eps_{i} B_i^2 e^{2v^i \tau} dy^i \otimes dy^i$$

     the parameters $v^i$ should obey the
     relations (\ref{3.20}):  $\sum_{i=1}^n  v^i = 0$ and
     \begin{equation}
        G_{i_1 ... i_{2m}}^{(2m)} v^{i_1} ... v^{i_{2m}} =
      (2m)! \sum_{i_1 <...< i_{2m}} v^{i_1} ... v^{i_{2m}} = 0.   \label{4.10}
      \end{equation}
    instead of (\ref{3.21}).

    The existence of these solutions corresponding to the
    ``pure''  Lowelock action (\ref{4.6})
    may be considered as  test for the validity of the conjecture suggested
    above.

 \section{Conclusions}

 Here we have considered the  $(n +1)$-dimensional  Einstein-Gauss-Bonnet
 model.
 For  diagonal  cosmological  type metrics we have reduced
 the equation of motion to a set of Lagrange equations with
  the Lagrangian   governed by two ``minisuperspace'' metrics on
 $\R^{n}$: (i) the pseudo-Euclidean 2-metric (corresponding to the scalar
 curvature term)  and  (ii) the Finslerian  4-metric
 (corresponding to the Gauss-Bonnet term).
 The  Finslerian  4-metric is proportional to
 $n$-dimensional  Berwald-Moor 4-metric. Thus, we have found a
 rather natural and  ``legitime'' application of
 $n$-dimensional Berwald-Moor  metric in multidimensional gravity
 with the Gauss-Bonnet term.  mth
  For the case of the ``pure'' Gauss-Bonnet model
  we have obtained  two  exact solutions:
  with power-law and exponential
  dependence of scale factors (w.r.t. ``synchronous-like'' variable).
  In the cosmological case (with $w =  -1$, $ \eps_{1}=  ... = \eps_{n} = 1$)
  the first (power-law) solution was obtained
  earlier by N. Deruelle for $n = 4, 5$ \cite{Deruelle}
  and verified by A. Toporensky and P. Tretyakov   (for $n = 6,7$) \cite{TT}
  and by S. Pavluchenko  (for all $n$) \cite{Pavl}. See also
  \cite{Iv-09}.

  When  the ``synchronous-like'' time
  gauge was considered   the equations of motion were reduced to
  an autonomous system of first order differential equations.
  It was shown that for any non-trivial  solution with
  the exponential dependence of scale factors $a_i(\tau) = A_i \exp( v^i
  \tau)$, $i = 1,...,n$, there are no more than
  three different  numbers among  $v^1,...,v^n$
  (if $\alpha_1 \neq 0$ and $\alpha_2 \neq 0$.).
  This   means  that the solutions of such type have a ``restricted''
  anisotropy. Such solutions  may be used for constructing
  of new  cosmological solutions, e.g. describing
  accelerated expansion of our 3-dimensional factor-space
  and small enough variation of the effective gravitational
  constant. For this approach, see  \cite{BZhuk,IKM-08}
  and references therein.

  We have also proposed (without a proof) a  generalization of the EGB
  effective \\
  (cosmological-type) Lagrangian  to the Lowelock  case
  (in agreement with  \cite{Deruelle,Pavl} for cosmological
  metrics).  According to this conjecture
  a ``pure'' Lowelock term of $m$-th order in the action gives a contribution
  to the  effective  Lagrangian  that
  contains a Finslerian  $2m$-metric. This hypothesis
  implies the existence of  cosmological solutions with power-law
  (see also \cite{Deruelle,Pavl} for cosmological case)
  and exponential  dependence of scale factors   for the case of the
  ``pure'' Lowelock model of $m$-th order.
   A proof of the conjecture mentioned above may be
   the subject of a separate publication. Another generalization of
   the approach suggested in this paper will be connected with
   inclusion of a scalar field.

  Here an open problem arises: do the
  generalized solutions (for arbitrary $n$)  with ``jumping''
  parameters $p^i, A_i$ appear as asymptotical  solutions in EGB
  model  when  approaching a singular  point?
  Recall that Kasner-type solutions with ``jumping''
  parameters $p^i, A_i$ describe an approaching to a singular
  point   in certain gravitational models, e.g. with
  matter sources, see
  \cite{BLK,DHSp,IKM-bil1,IKM-bil2,IM-bil1,IM-bil2,DamH1,DHN,IM-bil-rev}
  and references therein. This problem may be a subject of
  separate investigations. (Here it is worth to mention the
  paper of T. Damour and H. Nicolai \cite{DamNic}, which
  includes a study of the effect of the 4th order in curvature
  gravity terms, including the Euler-Lovelock term octic in
  velocities, and its compatibility with the Kac-Moody algebra
  $E_{10}$.)

 \begin{center}
 {\bf Acknowledgments}
 \end{center}

 This work was supported in part by the Russian Foundation for
 Basic Research grants Nr. 09-02-00677-a. The author is also
 grateful to A.V. Toporensky and D.G. Pavlov for  lectures
 at seminars of VNIIMS-RUDN,
 which stimulated  the writing of this paper. The
 main results of this work were reported at
 Vth International Conference
 ``Finsler Extensions of Relativity Theory'' (27 September -- 3
 October 2009, Moscow - Fryazino, Russia). The author thanks the
 participants of this conference for fruitful discussions
  and numerous comments.


\renewcommand{\theequation}{\Alph{subsection}.\arabic{equation}}
\renewcommand{\thesection}{}
\renewcommand{\thesubsection}{\Alph{subsection}}
\setcounter{section}{0}

\section{Appendix}

\subsection{Useful relations for $(1+n)$-splitting}

  Let us consider the metric defined on $\R_{*} \times \R^{n}$
  ($\R_{*} = (u_{-},u_{+})$ is an open subset in $\R$)
  \begin{equation}
   g= w e^{2{\gamma}(u)} du \otimes du  +
   \sum_{i,j =1}^{n} h_{ij}(u) dy^i \otimes dy^j. \label{A.1}
 \end{equation}

  Here $(h_{ij}(u))$ is a symmetric non-degenerate matrix for
  any $u \in \R_{*}$, smoothly dependent upon $u$.
  The function ${\gamma}(u)$ is smooth.

   The calculations give the following non-zero (identically)
   components of the Riemann tensor
   \begin{eqnarray}
    R_{0i0j} = - R_{i00j} = - R_{0ij0} = R_{i0j0} =
     \frac{1}{4} [-2 \ddot{h}_{ij} +
     2 \dot{\gamma} \dot{h}_{ij} +
     \dot{h}_{ik} h^{kl} \dot{h}_{lj}]  ,  \label{A.2}\\
    R_{ijkl} =  \frac{1}{4} (-w) e^{- 2 \gamma}
     (\dot{h}_{ik} \dot{h}_{jl} -
      \dot{h}_{il} \dot{h}_{jk}),  \label{A.3}
    \end{eqnarray}
   $i,j,k,l = 1, \dots, n$,
   where here and in what follows
   $h^{-1} = (h^{ij})$ is the matrix inverse to the matrix $h = (h_{ij})$.
   Here we denote  $\dot{A} = dA/du$ etc.

   For non-zero (identically)
   components of the Ricci tensor we get
   \begin{eqnarray}
    R_{00} =
     \frac{1}{2} [- h^{il} \ddot{h}_{li} +
     \frac{1}{2} h^{ij} \dot{h}_{jk}  h^{kl} \dot{h}_{li}+
    h^{ik} \dot{h}_{ki}  \dot{\gamma}],  \label{A.4}\\
    R_{ij} =  \frac{1}{4} (-w) e^{- 2 \gamma}
     [ 2 \ddot{h}_{ij} + \dot{h}_{ij}(h^{kl} \dot{h}_{lk} - 2
     \dot{\gamma}) -2 \dot{h}_{ik} h^{kl} \dot{h}_{lj} ],
     \label{A.5}
    \end{eqnarray}
    $i,j = 1, \dots, n$.

    The scalar curvature reads
     \begin{equation}
      R =  \frac{1}{4} (-w) e^{- 2 \gamma} [ 4 {\rm tr}(\ddot{h}h^{-1})
       + {\rm tr}(\dot{h}h^{-1}) ({\rm tr}(\dot{h}h^{-1}) - 4 \dot{\gamma})
       - 3  {\rm tr}(\dot{h} h^{-1}\dot{h}h^{-1})].   \label{A.6}
     \end{equation}

     Let us denote
     \begin{equation}
        M = \dot{h}h^{-1},         \label{A.8}
      \end{equation}
     ($h = (h_{ij})$), then
     \begin{equation}
     \dot{M} + M^2 = \ddot{h}h^{-1}. \label{A.10}
     \end{equation}

      We obtain
      \begin{equation}
      R \sqrt{|g|}=  L_1 + \frac{df_1}{du},   \label{A.7}
      \end{equation}
      where
      \begin{equation}
      L_1 =   \frac{1}{4} (-w) e^{- \gamma} \sqrt{|h|}
      [ {\rm tr}M^2 - ({\rm tr}M)^2 ],   \label{A.7L1}
      \end{equation}
        $|h| = |{\rm det}(h_{ij})|$ and
      \begin{equation}
        f_1 = (-w) e^{- \gamma} \sqrt{|h|} {\rm tr} M .    \label{A.7f1}
      \end{equation}

      In derivation of (\ref{A.7}) the following relations were
      used:
      \begin{equation}
        \frac{d\sqrt{|h|}}{du} = \frac{1}{2}\sqrt{|h|} {\rm tr}(\dot{h}h^{-1}),
         \qquad
        \sqrt{|g|} =  e^{\gamma} \sqrt{|h|}. \label{A.12}
       \end{equation}

     The calculations
     give us the following relations for quadratic invariants
     \begin{eqnarray}
       R_{MNPQ}  R^{MNPQ}   =  \frac{1}{8}  e^{- 4 \gamma}
      \{ ({\rm tr}M^2)^2 - {\rm tr}M^4 +
      2 {\rm tr}(2 \dot M + M^2 - 2 \dot{\gamma} M)^2 \},
      \label{A.13}\\
      R_{MN}  R^{MN} = \frac{1}{16}  e^{- 4 \gamma}
      \{ [ - 2 {\rm tr}\dot{M}  - {\rm tr} M^2  +
        2 \dot{\gamma}  {\rm tr} M ]^2 +
      {\rm tr}[2 \dot{M} + ({\rm tr} M - 2 \dot{\gamma})
      M]^2 \}. \label{A.14}
        \end{eqnarray}

    Relations (\ref{A.6}), (\ref{A.13}) and (\ref{A.14}) imply
    the following formula for the Gauss-Bonnet term (\ref{1.2})
   \begin{eqnarray}
    {\cal L}_2   = \frac{1}{16}  e^{- 4 \gamma}
    \{ 2 ({\rm tr}M^2)^2 - 2 {\rm tr}M^4 +
    [({\rm tr}M)^2 - {\rm tr}M^2] [8 {\rm tr} \dot{M}
    + 3 {\rm tr}M^2    \nonumber \\
     + ({\rm tr}M)^2 - 8 \dot{\gamma} {\rm tr} M ]
   + 4 {\rm tr}[( M^2 - ({\rm tr}M) M ) (4 \dot{M}
       -4 \dot{\gamma} M  + M^2 + ({\rm tr}M)M)] \}.
   \label{A.15}
   \end{eqnarray}

   Relation  (\ref{A.15}) implies another important formula
     \begin{equation}
        {\cal L}_2  \sqrt{|g|} = L_2 + \frac{d}{du} f_2, \label{A.16}
     \end{equation}
    where
    \begin{eqnarray}
       L_2  = \frac{1}{48}  e^{- 3 \gamma} \sqrt{|h|}
       \{ 6 {\rm tr}M^4 - 3 ({\rm tr}M^2)^2
       \nonumber \\
       + 6 {\rm tr}M^2 ({\rm tr}M)^2 - 8 ({\rm tr}M) {\rm tr}M^3
       - {\rm tr}M^4 \}
       \label{A.16L2}
    \end{eqnarray}
   and
    \begin{equation}
     f_2 = \frac{1}{6}  e^{- 3 \gamma} \sqrt{|h|}
     \{ 2 {\rm tr}M^3 - 3 ({\rm tr}M) {\rm tr}M^2
       +    ({\rm tr}M)^3 \}.
       \label{A.16f2}
   \end{equation}

{\bf Diagonal metrics.}

 Now we consider the diagonal metric
    \begin{equation}
    h_{ij}(u) = e^{2\beta^i(u)} \eps_i \delta_{ij},  \label{A.17h}
    \end{equation}
 $\eps_i = \pm 1$, $i = 1, \ldots, n$.
 Then, $M_{ij} = 2 \dot{\beta}^i \delta_{ij}$
 and we get the following relations for ``Lagrangians''
   \begin{eqnarray}
       L_1 = (-w) e^{-\gamma + \gamma_0}
       \left[\sum_{i =1}^{n} (\dot{\beta}^i)^2 -
        (\sum_{i =1}^{n} \dot{\beta}^i)^2 \right]
         \label{A.18L1}   \\
       L_2  = - \frac{1}{3}  e^{- 3 \gamma + \gamma_0}
       \left\{ (\sum_{i =1}^{n} \dot{\beta}^i)^4
        - 6 (\sum_{i =1}^{n} \dot{\beta}^i)^2 \sum_{j =1}^{n}
        (\dot{\beta}^j)^2  \right.
         \nonumber \\
     \left.   +     3  (\sum_{i =1}^{n} (\dot{\beta}^i)^2)^2
        + 8  (\sum_{i =1}^{n} \dot{\beta}^i) \sum_{j =1}^{n} (\dot{\beta}^j)^3
        - 6 \sum_{i =1}^{n} (\dot{\beta}^i)^4 \right\},
       \label{A.18L2}
    \end{eqnarray}
  where $\gamma_0 = \sum_{i =1}^{n} \beta^i$.

  The ``f-functions'' (\ref{A.7f1}) and (\ref{A.16f2}) read
  as follows
    \begin{eqnarray}
       f_1 = 2 (-w) e^{-\gamma + \gamma_0}
       \sum_{i =1}^{n} \dot{\beta}^i,
                \label{A.18f1}   \\
       f_2  = \frac{4}{3}  e^{- 3 \gamma + \gamma_0}
       \left[ 2 \sum_{i =1}^{n} (\dot{\beta}^i)^3
        - 3 (\sum_{i =1}^{n}  \dot{\beta}^i) \sum_{j =1}^{n} (\dot{\beta}^j)^2
        + ( \sum_{i =1}^{n} \dot{\beta}^i)^3 \right].
       \label{A.18f2}
    \end{eqnarray}

   \addtocounter{section}{1} \setcounter{equation}{0}
  \subsection{Useful relations for Finslerian 4-metric}

 Here we consider a  proof of identity (\ref{2.13}).
 We decompose the product of $6$ terms in the definition
 of the 4-metric (\ref{2.10}) into the sum
 (of ``powers of $\delta$-s'')
 \begin{equation}
   G_{ijkl} = \sum_{a= 0}^{6} G_{ijkl}^{a} \label{B.1}
 \end{equation}
 where
   $$G_{ijkl}^{0} = 1,  \quad
  G_{ijkl}^{1} = - \delta_{ij} - \delta_{ik}  - \delta_{il}
                  - \delta_{jk} - \delta_{jl} -  \delta_{kl},
       ...,
  G_{ijkl}^{6} =  \delta_{ij}  \delta_{ik}  \delta_{il}
                    \delta_{jk}  \delta_{jl}  \delta_{kl}.$$

  Then we get
   \begin{equation}
   T = G_{ijkl}v^i v^j v^k v^l = \sum_{a= 0}^{6} T^a,  \label{B.2}
   \end{equation}
  where $T^a = G_{ijkl}^{a} v^i v^j v^k v^l$.

   The calculations of $T^a$ give us the following results:
      \begin{eqnarray}
      T^0 = S_1^4, \quad T^1 = - 6  S_1^2 S_2, \quad T^2 = 3 S_2^2 + 12  S_1 S_3,
      \nonumber \\
      T^3 = - 4  S_1 S_3 - 16 S_4, \quad T^4 = 15 S_4, \quad T^5 = - 6 S_4,
       \quad T^6 =  S_4, \label{B.3}
       \end{eqnarray}
      where
      \begin{equation}
       S_k = S_k (v) = \sum_{i =1}^n (v^i)^k, \label{B.4}
       \end{equation}
       $k = 1,2,3,4$.

    The summation of all  $T^a$ in (\ref{B.3})
    leads us to the relation
        \begin{equation}
    T = G_{ijkl}v^i v^j v^k v^l =
     S_1^4 -  6  S_1^2 S_2 + 3 S_2^2 + 8  S_1 S_3 - 6 S_4
      \label{B.17}
    \end{equation}
    coinciding with  (\ref{2.13}).

    Now we  prove relation  (\ref{5.4}).
   We get
   \begin{equation}
   P_i = G_{ijkl} v^j v^k v^l = \sum_{a= 0}^{6}P^a_i,  \label{B.18}
   \end{equation}
  where $P^a_i = G_{ijkl}^{a}  v^j v^k v^l$, $i = 1, \dots, n$.

   The calculations of $P^a_i$ give us the following formulas
      \begin{eqnarray}
      P^0_i = S_1^3, \quad P^1_i = - 3 S_1^2 v^i - 3  S_1 S_2,
      \quad P^2_i = 3S_3 + 3S_2 v^i + 9 S_1 (v^i)^2,
      \nonumber \\
        P^3_i = - S_3 - 3S_1 (v^i)^2  - 16 (v^i)^3 , \quad P^4_i = 15 (v^i)^3,
       \quad P^5_i = - 6 (v^i)^3,
       \quad P^6_i = (v^i)^3,       \label{B.19}
       \end{eqnarray}
      $i = 1, \dots, n$.

    The summation of all  $P^a_i$ in (\ref{B.19})
    leads us to the relation
       \begin{equation}
       P_i = S_1^3  + 2 S_3 - 3 S_1 S_2  + 3(S_2 - S_1^2) v^i
                   + 6 S_1 (v^i)^2 - 6(v^i)^3,
        \label{B.20}
       \end{equation}
    $i = 1, \dots, n$, coinciding with  (\ref{5.4}). This relation
    implies $P_i v^i = T$ in agreement with the
    definitions (\ref{B.2}) and (\ref{B.18}).

  \addtocounter{section}{1} \setcounter{equation}{0}
  \subsection{Lagrange equations}

  Here we prove the relations  (\ref{2.16a})-(\ref{2.16c})
  for the cosmological type metric  (\ref{2.2})
  defined on manifold $M$ from (\ref{2.1}).
  The tensor ${\cal E}_{MN}$ is obtained from
  the variation of the action
     \begin{equation}
   S =  \int_{M} d^{D}z \sqrt{|g|} {\cal L}[g],
     \label{C.1}
   \end{equation}
  with ${\cal L}[g] = \alpha_1 R[g]
       + \alpha_2  {\cal L}_2[g]$, i.e.
    \begin{equation}
   \delta S =  \int_{M} d^{D}z \sqrt{|g|} {\cal E}_{MN} \delta g^{MN},
     \label{C.2}
   \end{equation}
   and $\sqrt{|g|} {\cal E}_{MN} = \delta S/ \delta g^{MN}$.

   Without loss of generality any 1-dimensional submanifold $M_i$
   is chosen to be compact and coinciding with the circle of unit length: $M_i = S^1_r$
   ($r = 1/2 \pi$ is the radius of the circle) and all
    coordinates $y^i$ (see (\ref{2.1})) obey $0 < y^i <1$,
    $i = 1, \dots,n$.

    Here we will use the following relations
    for the components  ${\cal E}_{MN}$
    in coordinates $(y^M) = (y^0 =u, y^i$)
    and   ${\cal L}$   calculated for the metric (\ref{2.2}):
      \begin{eqnarray}
     {\cal E}_{MN} = \delta_{MN} {\cal E}_{NN}, \label{C.3} \\
     {\cal E}_{MN} = {\cal E}_{MN}(u),    \label{C.4} \\
     \sqrt{|g|}  {\cal L} = L + \frac{df}{du},
     \label{C.5}
     \end{eqnarray}
 where  $L = L(\gamma, \beta, \dot{\beta})$
 and  $f = f(\gamma, \beta, \dot{\beta})$ are defined
 in relations  (\ref{2.4}) and (\ref{2.7}), respectively.

 The first relation (\ref{C.3}) may be readily verified using
  (\ref{1.3e})-(\ref{1.3b}) and formulas
  for the Riemann tensor (\ref{A.2}) and (\ref{A.3}). The second
 relation  (\ref{C.4}) is an obvious one and the third one
 (\ref{C.5})  is coinciding  with (\ref{2.3}).

 The substitution of the metric (\ref{2.2}) into the functional (\ref{C.1})
 gives us (due to (\ref{C.5}) and $0 < y^i <1$)
   \begin{equation}
   S =  \int_{u_{-}}^{u_{+}} du \left (L + \frac{df}{du} \right)
     \label{C.6}
   \end{equation}
 and hence
   \begin{equation}
   \delta S =  \int_{u_{-}}^{u_{+}} du \left\{
    \frac{\p L}{\p \gamma}  \delta \gamma
     + \sum_{i=1}^n \left( \frac{\p L}{\p \beta^i} -
       \frac{d}{du} \frac{\p L}{\p \dot{\beta}^i}\right)
    \delta \beta^i \right\}, \label{C.7}
   \end{equation}
    where $\delta \gamma(u)$ and  $\delta \beta^i(u)$ are smooth
   functions with compact support in $(u_{-}, u_{+})$
   ($\delta \gamma(u_{\pm}) = \delta \beta^i(u_{\pm}) = 0)$,
    $i = 1, \dots,n$. On the other hand, using
    (\ref{C.2})-(\ref{C.4}), the relation
    $$(\delta g^{MN}) = {\rm diag} (-2w e^{- 2 \gamma} \delta \gamma,
     - 2 \eps_1 e^{- 2 \beta^1}\delta \beta^1, \dots, - 2 \eps_n e^{- 2
    \beta^n}\delta \beta^n)$$
    and $0 < y^i <1$, we get
   \begin{equation}
   \delta S =  \int_{u_{-}}^{u_{+}} du
  \{ {\cal E}_{00}(-2w) e^{\gamma_0 - \gamma} \delta \gamma +
   \sum_{i=1}^n  {\cal E}_{ii}(-2 \eps_{i})
   e^{\gamma + \gamma_0 - 2\beta^i} \delta \beta^i \}. \label{C.8}
        \end{equation}

  Comparing  (\ref{C.7}) and (\ref{C.8}) we get
  relations (\ref{2.16a}) and (\ref{2.16b}). Relations  (\ref{2.16c})
  just follow from  (\ref{C.3}).

  \addtocounter{section}{1} \setcounter{equation}{0}
  \subsection{Riemann tensor squared}

  Here we consider the Riemann tensor squared (Kretchmnann scalar) for the metric
  (\ref{3.12})

    $$g= w d \tau \otimes d \tau  +
  \sum_{i=1}^{n} \eps_{i} A_i^2 \tau^{2p^i} dy^i \otimes dy^i.$$

   From (\ref{A.13}) we get
   \begin{equation}
      R_{MNPQ}  R^{MNPQ}  = K \tau^{-4},        \label{D.2}
   \end{equation}
   where
    \begin{equation}
      K  = 2 S_4 + 2 S_2^2 - 8 S_3 + 4 S_2        \label{D.3}
   \end{equation}
   and $S_k = S_k (p) = \sum_{i =1}^n (p^i)^k$, $k = 1,2,3,4$.

   Using the identities
   \begin{equation}
      K  = 4 \sum_{i = 1}^n (p^i - 1)^2 (p^i)^2 + 2 (S_2^2 - S_4).       \label{D.4}
   \end{equation}
   and
    \begin{equation}
       S_2^2 - S_4 = 2 \sum_{i < j} (p^i)^2 (p^j)^2       \label{D.5}
   \end{equation}
   we obtain that $K \geq 0$ and $K = 0$ if and only if
   the set of   parameters $p = (p^1,...,p^n)$
   is either trivial: $p = (0,...,0)$, or
   belongs to the Milne set:
      \begin{equation}
      p = (1,0,...,0), \ldots,    (0,...,0,1).            \label{D.6}
      \end{equation}

   For other sets $p$ we have $K > 0$ and the Riemann tensor squared
    diverges when $\tau \to + 0$.

\addtocounter{section}{1} \setcounter{equation}{0}
  \subsection{The proof of Proposition 1}

         The equations of motion  (\ref{5.1}) and (\ref{5.2})
         corresponding to the metric  (\ref{3.12}) with $h^i = p^i/\tau$
         (here $\alpha_1 =0$  and $\alpha_2 \neq 0$)  read as
         follows
         \begin{eqnarray}
       {\cal A} \equiv  G_{ijkl}p^i p^j p^k p^l = 0,
              \label{E.1}     \\
        {\cal D}_i \equiv G_{ijkl} p^j p^k p^l  = 0,
                   \label{E.2}
       \end{eqnarray}
       $i = 1,\dots, n$.

        Let $D = n+1 \neq 4$  and
         \begin{eqnarray}
       {\cal B} \equiv  \frac{1}{(n - 3)} \sum_{i =1}^{n} {\cal D}_i,
            \label{E.6} \\
      {\cal C}_i \equiv \frac{1}{3} ({\cal B} - {\cal D}_i) \label{E.7},
        \end{eqnarray}
        $i = 1,\dots, n$.

        For $D \neq 4$ the set of  equations (\ref{E.1}) and (\ref{E.2})
        is equivalent to the following set of equations
        \begin{eqnarray}
    {\cal A} =  S_1^4 -  6  S_1^2 S_2 + 3 S_2^2 + 8  S_1 S_3 - 6 S_4 =
      24 \sum_{i < j < k < l} p^i p^j p^k p^l = 0,
        \label{E.10}     \\
    {\cal B} =  (S_1 - 3)(S_1^3  - 3 S_1 S_2 + 2 S_3) =
                  6(S_1 - 3)\sum_{i < j < k } p^i p^j p^k = 0,
            \label{E.11} \\
    {\cal C}_i =  (S_1 - 3) p^i [2 (p^i)^2 - 2 S_1 p^i + S_1^2 - S_2 ] = 0,
                   \label{E.12}
      \end{eqnarray}
      $i = 1,\dots, n$.
      Here $S_k = S_k (p) = \sum_{i =1}^n (p^i)^k$ and
      we  used the identities (\ref{2.13}),
        (\ref{5.4}) and the following identity
      \begin{equation}
     S_1^3  - 3 S_1 S_2 + 2 S_3 = G_{ijk}p^i p^j p^k = 6 \sum_{i < j < k } p^i p^j  p^k,
     \label{E.8}
      \end{equation}
      where
        \begin{equation}
        G_{ijk}  = (\delta_{ij} -1)(\delta_{ik} -1)(\delta_{jk} -1)
        \label{E.9}
        \end{equation}
      are  components of  a  Finslerian 3-metric.
      The identity  (\ref{E.8}) could be readily verified
      along a line as it was done in Appendix B for the Finslerian 4-metric.
      (We note that relation (\ref{E.11}) may be also obtained
      using the formula (\ref{A.16}).)

      For $S_1 = 3$ we obtain the main solution governed
      by relations (\ref{3.13}) and (\ref{3.14}).

       Now we consider another case $S_1 \neq 3$. Let $k$
      be the number of all nonzero numbers  among $p^1,...,p^n$.
      For $k = 0$ we get a trivial solution $(0,...,0)$.
      Let $k \geq 1$. We suppose without loss of generality that
      $p^1,...,p^k$ are nonzero. For $k = 1, 2$ all relations
      (\ref{E.10})-(\ref{E.12}) are satisfied identically.
      In all three cases $k = 0, 1, 2$  the solutions have
      the form $(a,b,0..,0)$ (plus permutations for general setup).

       Now we consider $k \geq 3$. From (\ref{E.12}) and
       $S_1 \neq 3$  we obtain
        \begin{equation}
         2 (p^i)^2 - 2 S_1 p^i + S_1^2 - S_2 = 0,
          \label{E.13}
        \end{equation}
        $i = 1,\dots, k$. Summing  on $i$ gives us $(2 -k) (S_2 - S_1^2) =
          0$, or $S_2 = S_1^2$. Then we obtain from
        (\ref{E.11}) $S_3 = S_1^3$ and from
       (\ref{E.10}): $S_4 = S_1^4$. Thus, we get
        $S_4 = S_2^2$ implying $\Sigma = \sum_{1 \leq i < j \leq k } (p^i)^2 (p^j)^2 =
        0$. But  $\Sigma \geq (p^1)^2 (p^2)^2 > 0$. Hence, we
          are led to a contradiction. That means that
         for $S_1 \neq 3$, we have only  solutions with $k \leq 2$
         of the form $(a,b,0..,0)$ (plus permutations for general setup).
         The Proposition 1   is proved.

\end{document}